\documentclass[article,showpacs]{revtex4}
\renewcommand{\baselinestretch}{2}
\usepackage{graphicx}
\usepackage{dcolumn}
\usepackage{bm}
\begin{document}
\title{The Network of Commuters in London}
\author{A. P. Masucci}
\author{G. J. Rodgers}
\affiliation{%
Department of Mathematical Sciences, Brunel University, Uxbridge,
Middlesex, UB8 3PH, United Kingdom}%

\date{\today}
\begin{abstract}
We study the directed and weighted network in which the wards of
London are vertices and two vertices are connected whenever there is
at least one person commuting to work from a ward to another.
Remarkably the in-strength and in-degree distribution tail is a
power law with exponent around $-2$, while the out-strength and
out-degree distribution tail is exponential. We propose a simple
square lattice model  to explain  the observed empirical behaviour.
\end{abstract}
 \maketitle

\section{\label{sec:level1}Introduction.\protect}

Applications of graph theory to the study of urban development has a
long history, initiated by Euler's study of urban traffic problems
\cite{e1,e2}. A review of the  state of the art on cities and
complexity, studied  through cellular automata, agent-based models
and fractals can be found in \cite{b}. Following the seminal work of
Barabasi et al. on growing scale free networks \cite{4}, many
attempts have been made to embed growing networks in a Euclidean
bi-dimensional space\cite{h}. Some of these consider vertices to be
random points in a selected space\cite{m}, some others consider
vertices to be cells in a given lattice\cite{r,y}. Moreover spectral
analysis on urban networks \cite{v} have shown to unreveal many
interesting aspects of metropolitan organisation.

  We study the network of commuters in London. Data on commuters' behaviour
  was obtained from the London 2001 census that is available in \cite{lc}.
London is composed of 634 wards. We  consider the network in which
vertices are wards and two vertices are linked whenever there is a
flux of people commuting  from a ward to another to work. Loops are
considered. Since this network is embedded in a geographical space
we
 need to introduce a definition of physical distance between the
 vertices. In a city like London the Euclidean distance is not the
 best choice if we want to deal with the organization of people
  and the development of the city. Many places in the city can be very close in terms of
 Euclidean distance, but far apart if we consider their accessibility,
 that is the time a person would take to  commute from one place to another.
 For this reason we adopt as a distance between two wards the
 $generalised$ $time$ $t^*$ to travel from a ward to the other. The definition and data sets for the  generalised time were
 developed by Transport for London \cite{tfl} for the Greater London Authority.   The generalised time
 is defined as $t^* =in-vehicle$ $time+2*(waiting$ $time) +1.5*(access$ $ and$ $ egress$ $time)
 + interchange +$ $bus$ $boarding$ $penalty$ and it is measured in
 minutes.

 All data available concern just London. For instance people living
out of London and working in London, or people living in London and
working out of London are not counted. This bias can be important if
we consider that the 25$\%$ of people working in the central
activity zones of London live out of London.

The links of this network are directed and weighted. The
directionality of the network is implicit in the complexity of urban
commuting. The city is composed of wards that are mainly devoted to
business, wards that are mainly residential  and wards that are both
business and residential oriented. This implies the way people
commute from a ward to another is strongly ward dependent and
directional. We will consider people out-going from the ward where
they live and in-coming to the ward where they work. The  result of
this approach is that the in and  out vertices properties are
different for different wards and give light to two different
mechanisms involved in the development of the city. A weighted
analysis of this network is motivated by the fact that the flux of
people commuting from one ward to another is an important measure of
the dynamics of the city.

We   define the weighted adjacency matrix $W=\{w_{ij}\}$ , $i,
j=1,2,...,634$, for the network, where $w_{ij}$ is the weight of the
link connecting the vertex $i$ to the vertex $j$, that is the number
of people  living in ward $i$ and commuting to  ward $j$ to work.
Note that, since the network is directed, this number will be
different from $w_{ji}$, that is the number of  people living in
ward $j$ and working in ward $i$, i.e. the matrix is not symmetric.
We define the out and in-degree $k_i^{out/in}$ of a vertex $i$ as
the number of its first out/in nearest neighbours, that is  ,
$k_i^{out/in}=\sum_{j=1}^{634}\Theta(w_{ij/ji}-\epsilon)$, where
$\epsilon=o(w_{ij})$ and $\Theta$ is the Heaviside function  defined
as: $\Theta(x)=0$ if $x<0$, $\Theta(x)=1$ if $x>0$. The out-degree
of the vertex $i$ represents the number of different wards people
who live in ward $i$  work in. The in-degree of the vertex $i$
represents the number of different wards people who work in ward $i$
live in. We define the out/in-strength $s_i^{out/in}$ of a vertex
$i$ as the total out/in number of commuters, departing from/going to
the ward $i$, that is $s_i^{out/in}=\sum_{j=1}^{634}w_{ij/ji}$.

\begin{figure}[!htbc]\center
         \includegraphics[width=0.48\textwidth]{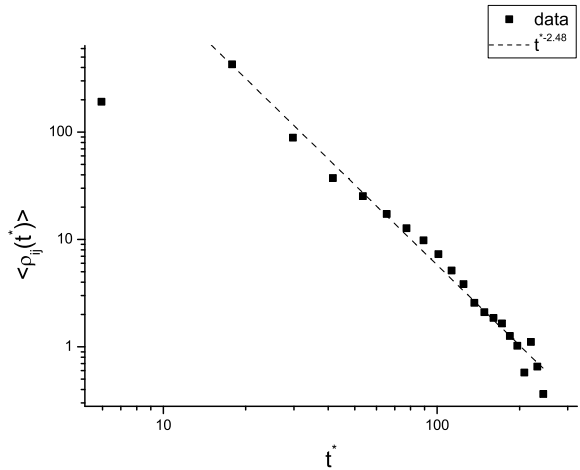}
\caption{\label{1} Average density of people $<\rho_{ij}>$ commuting
from ward $i$ to ward $j$ as a function of the generalised time $t^*$ to
travel from ward $i$ to ward $j$. The power law behaviour is a
signature of the strong geographical dependence of the network. }
\end{figure}

Since the  quantities  defined above are dependent on the size of
the wards, in order  to describe our system we will consider the
strength and degree area densities, both measured in $km^{-2}$. We
first define the weighted adjacency matrix $R=\{\rho_{ij}\}$ where
$\rho_{ij}=\frac{w_{ij}}{A_i}$ and $A_i$ is the area of ward $i$
measured in $km^2$. $\rho_{ij}$ represents the density of commuters
moving from ward $i$ to ward $j$. Our decision to use a real density
as the standard quantities to analyse our system is supported by the
fact that $\rho_{ij}$ shows a strong dependence on $t^*$, that is
$\rho_{ij}(t^*) \propto {t^*}^{-2.48}$ (see Fig.\ref{1}). This power
law behaviour demonstrates the strong geographical dependence of the
network.

We then define the degree density  $\tau_i^{out/in}$ for ward $i$ as
$\tau_i^{out/in}=\frac{k_i^{out/in}}{A_i}$ and the strength density
$\sigma_i^{out/in}$ for ward $i$ as
$\sigma_i^{out/in}=\frac{s_i^{out/in}}{A_i}$.

In section II we will show the main results of the empirical
analysis of the data. In section III we will propose a simple model
to reproduce the behaviour of commuters in the city.

\section{Empirical analysis}
The network is composed by 634 vertices connected by 143102 edges
with an average degree $<k>=226$, so that we can say it is a very
well connected network indeed.

\begin{figure}[!htbc]\center
\includegraphics[width=0.52\textwidth]{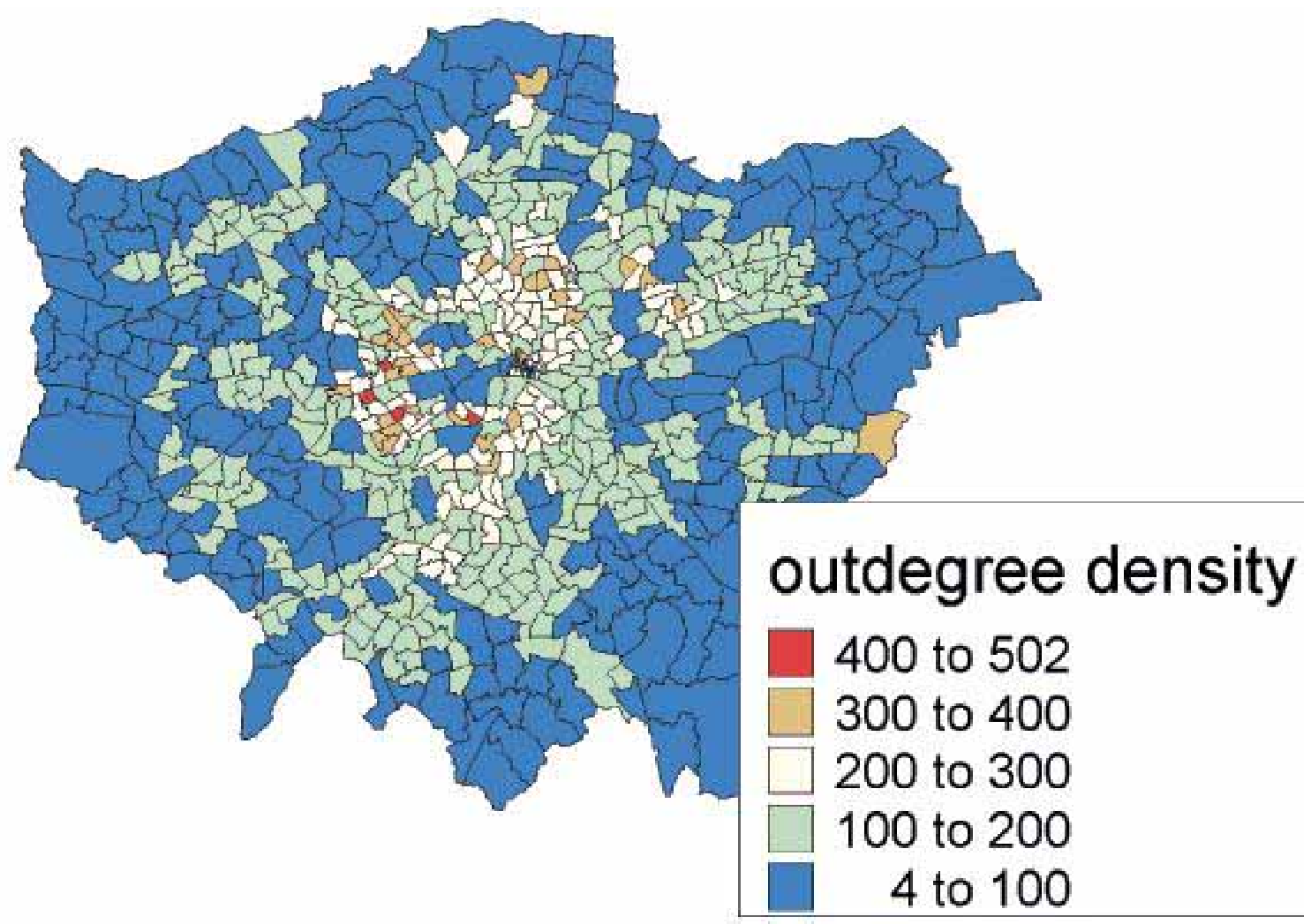}
             \includegraphics[width=0.47\textwidth]{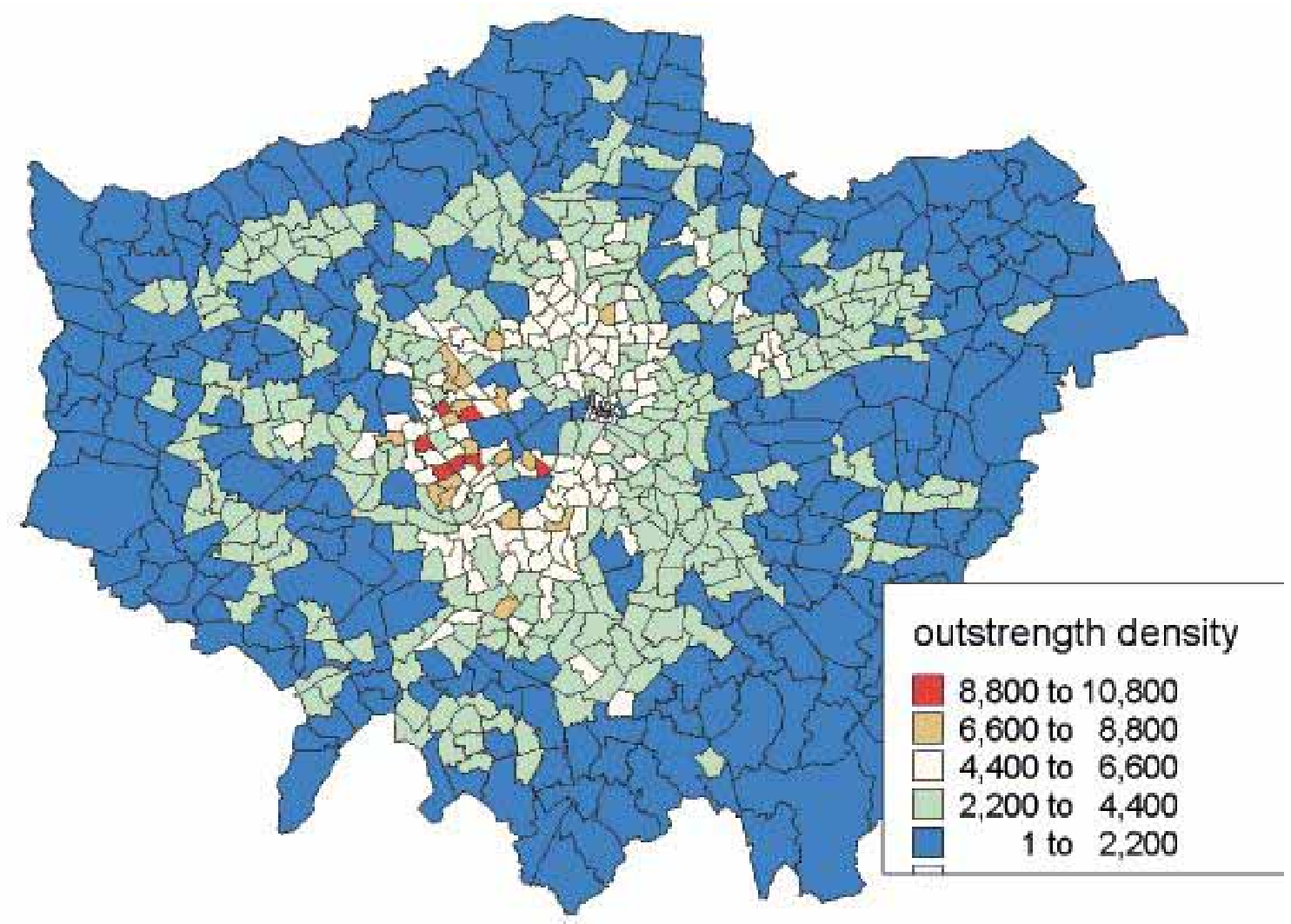}
             \includegraphics[width=0.52\textwidth]{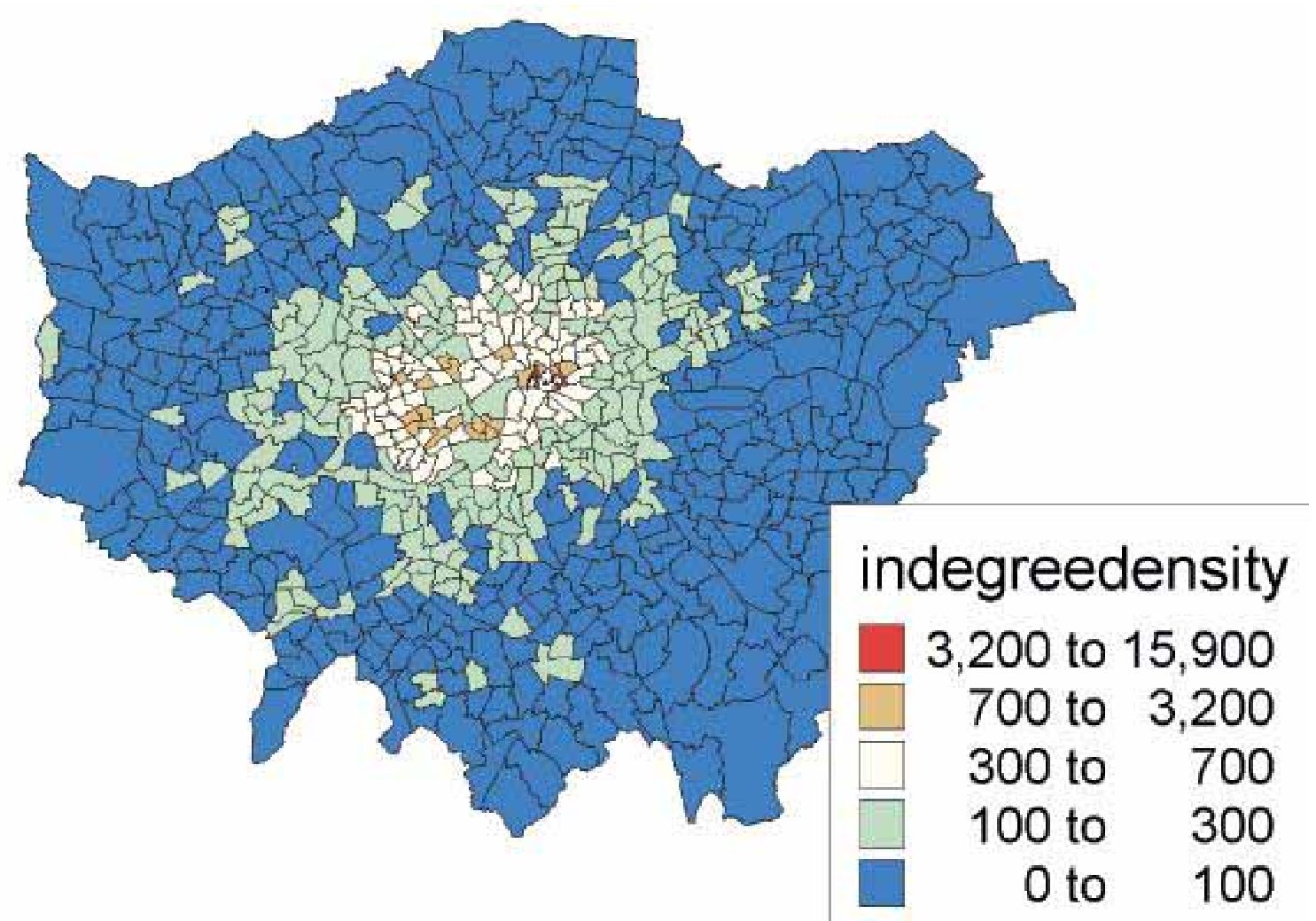}
             \includegraphics[width=0.45\textwidth]{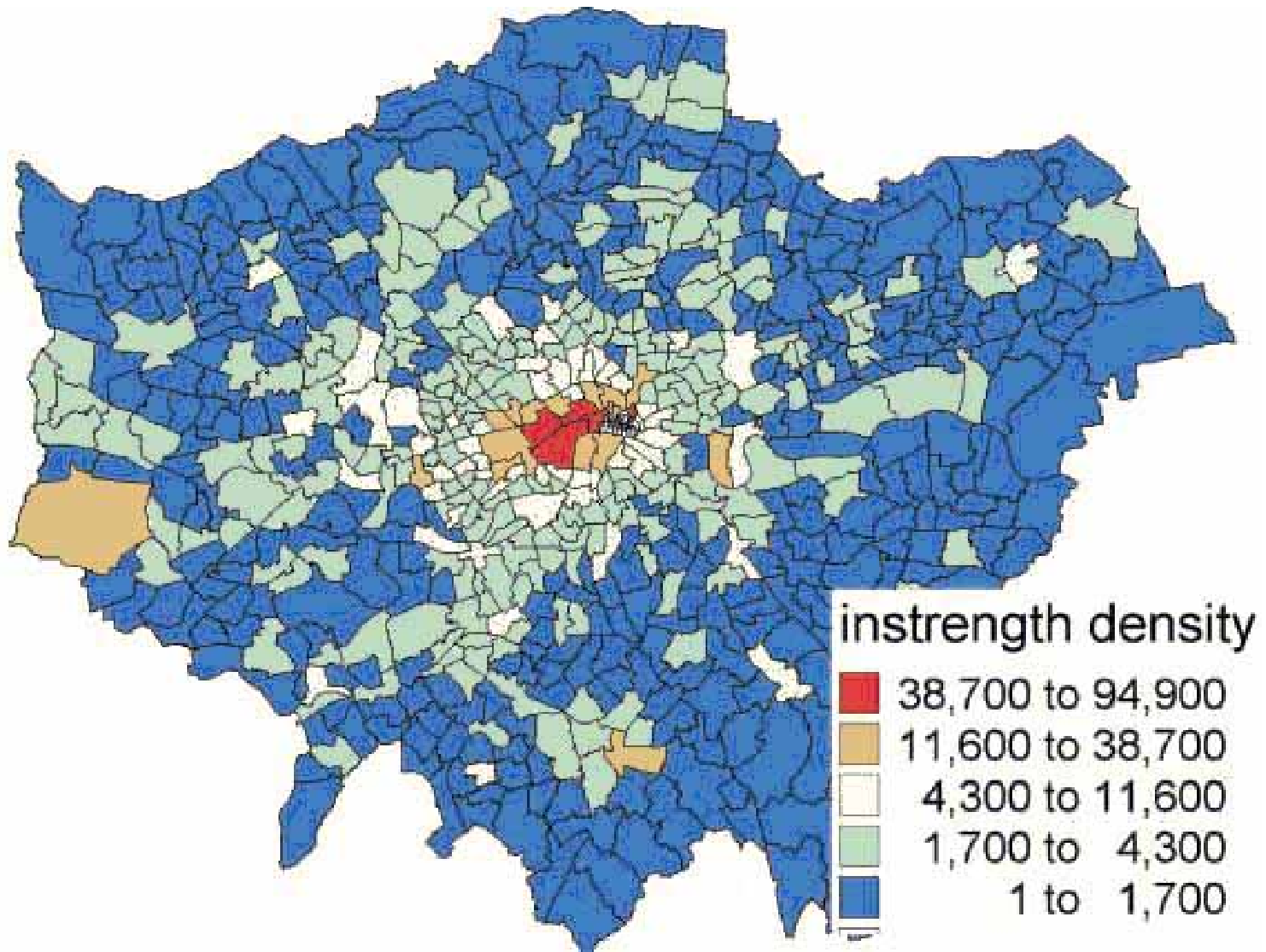}

\caption{\label{2} Maps of London showing the geographical
distribution for:  top left the out-degree density $\tau_{out}$; top
right the out-strength density $\sigma_{out}$; bottom left the
in-degree density $\tau_{in}$; bottom right the in-strength density
$\sigma_{in}$. }
\end{figure}

The out-degree density or out-connectivity density $\tau_i^{out}$ of
the vertex $i$ is the number of different wards people living in the
ward $i$  work in, divided for the area of the ward $i$, that is the
area density of working connections a ward can create with other
wards in London. It can be seen as a measure of the average
commuting chances of a ward. The out-degree density, in this
network, spans from values of 4.9, for $Darwin$ ward, to 501, for
$Aldersgate$ ward. The average out-degree density is 142.8. Examples
of wards with out-degree density around the average are: $St.
Pancras$ $and$ $Somers$ $Town$ ward, $The$ $Wrythe$
 ward,  etc.. In the top left of Fig.\ref{2} the map
of the geographical distribution of the wards out-degree density is
shown. It is interesting to notice how wards are organized in zones
well defined within this measure. In particular the light coloured
ring around the center in Fig.\ref{2} looks to be a good zone to
commute to any other area of London.

The in-degree density or in-connectivity density $\tau_i^{in}$ of
the vertex $i$ represents the number of wards  people  working in
ward $i$ live in, divided for the area of ward $i$. It can be seen
as a measure of the accessibility of a ward in respect of the other
wards. The in-degree density, in this network, spans from values of
2, for $Darwin$ ward, to 15525, for $Walbrook$ ward. The average
in-degree density is around 226. Examples of wards with in-degree
density around the average are: $Canonbury$ ward, $Southfield$ ward,
etc.. In the bottom left of Fig.\ref{2} it is shown the map of the
geographical distribution of the wards' in-degree density. From
those maps it is possible to appreciate how this measure can
identify the major business areas of London, where the in-degree is
large, and the residential areas where the in-degree is small.

From these first results we can observe that the in-degree has a
range of values that is larger by two orders of magnitude than that
of the out-degree. This result reflects two very different phenomena
behind the distribution process for settlement and  business areas.
This is due to the wards selectivity for urban
  function, where business tends to be concentrated in
  few areas while residential wards tend to spread over a much broader  region.

\begin{figure}[!htbc]\center
             \includegraphics[width=0.48\textwidth]{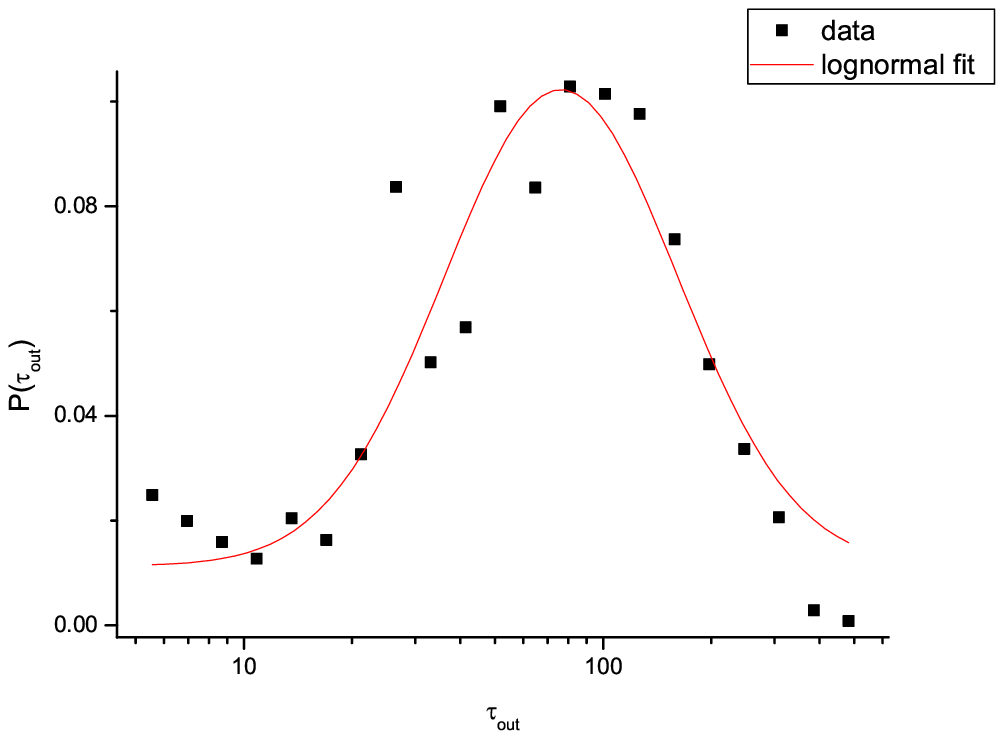}
             \includegraphics[width=0.48\textwidth]{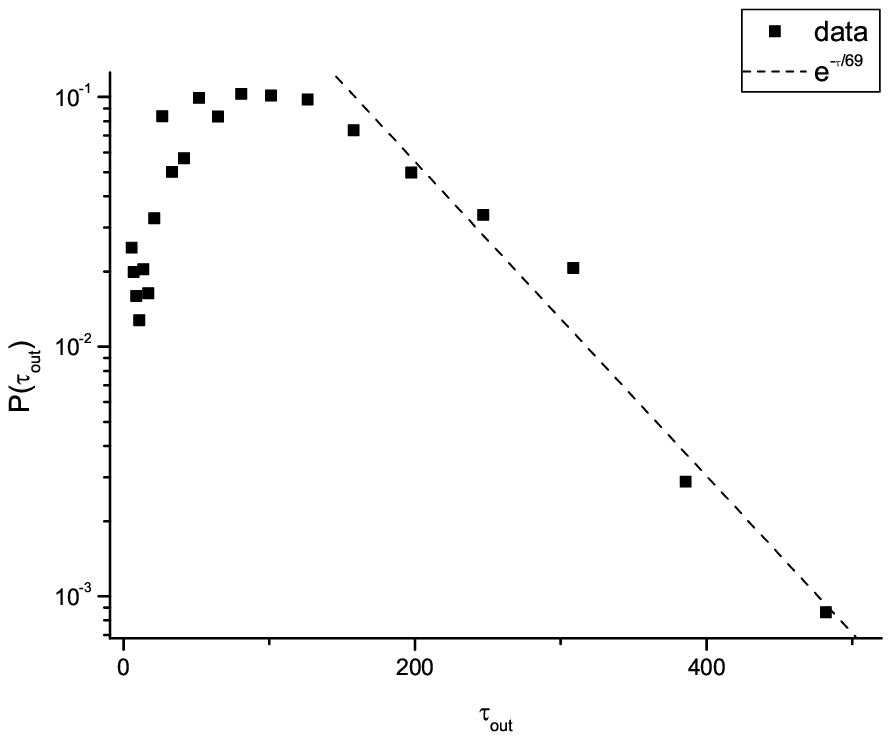}
             \includegraphics[width=0.48\textwidth]{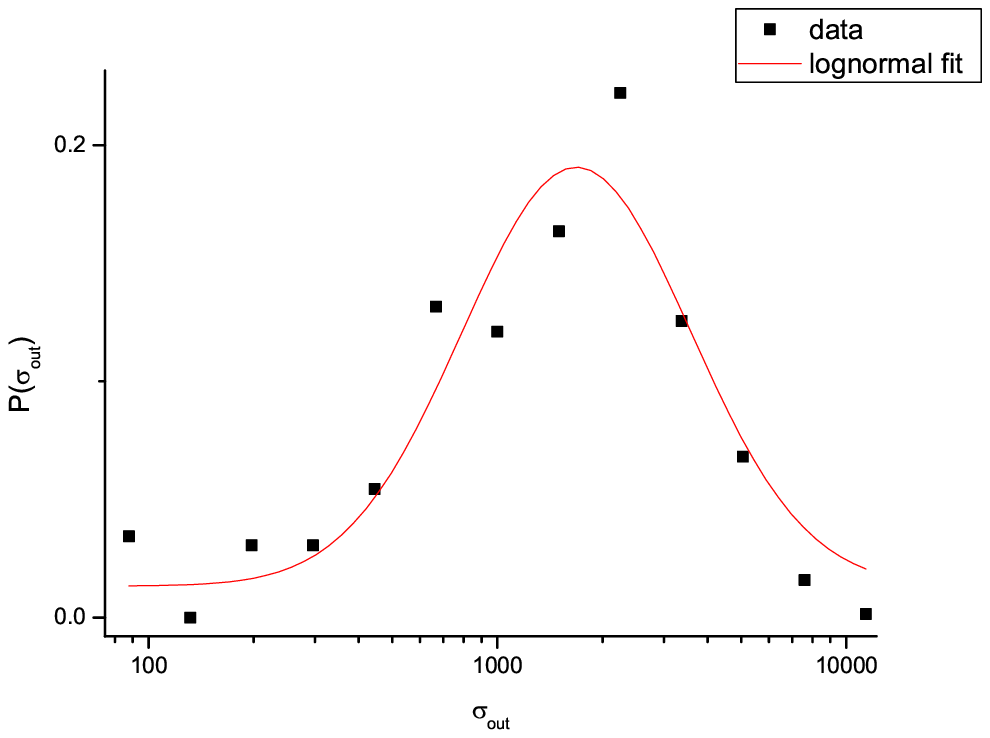}
             \includegraphics[width=0.48\textwidth]{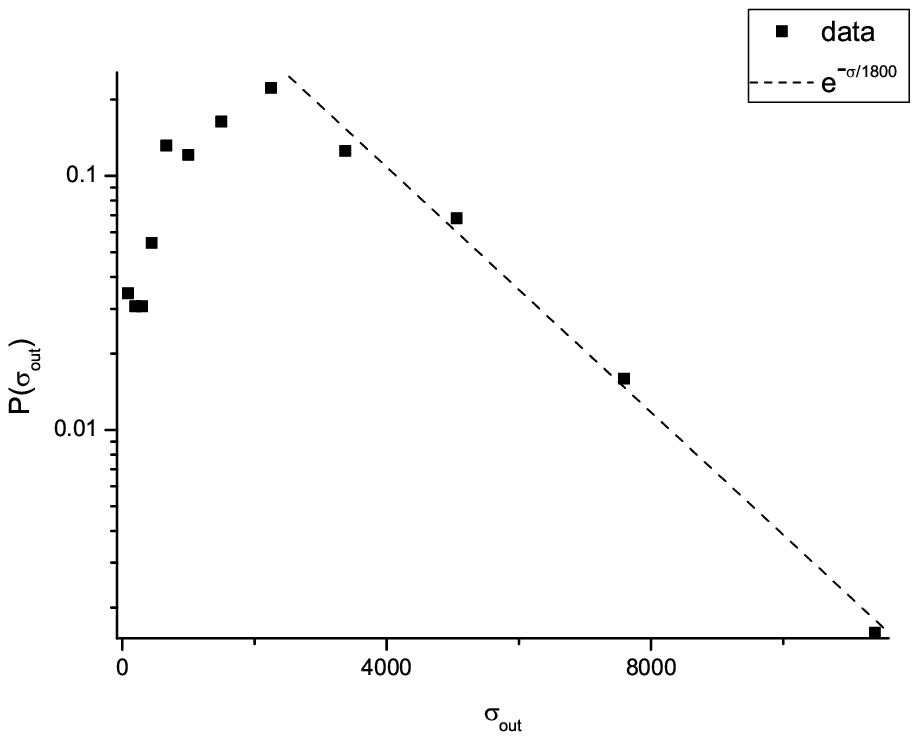}
\caption{\label{3}Top panels: Distribution for the out-degree
density $\tau_{out}$. On the left the log-linear scale shows a log-normal shape, while the log-log scale on the right shows the
power law behaviour in the tale of the distribution. Bottom
panels:Distribution for the out-strength density $\sigma_{out}$. On
the left the log-linear scale shows the log-normal shape, while
the log-log scale on the right illustrates the power law behavior of
the tale of the distribution. }
\end{figure}
The out-strength density $\sigma^{out}_i$  of the vertex $i$
represents the area density of employed people living in the ward
$i$. The out-strength density values in our data span from a minimum
of 70.20 for $Darwin$ ward, to a maximum of 10771.70 for $Earl's$
$Court$, with an average of 3051.38. Wards like $Golborne$,
$Nunhead$, $Muswell$ $Hill$, etc… are around the out-strength
average values. The geographical distribution for the wards
out-strength is given in the map in the upper right  of Fig.\ref{2}.

The in-strength density  $\sigma^{in}_i$ of a vertex $i$ represents
the total area density of people living in London who work in  ward
$i$, including  ward $i$ itself. The in-strength density can be seen
as a measure of the business capacity of a ward. The in-strength
density values span from a minimum of 158.48 for $Darwin$ ward, to a
maximum of 94805.70 for $St. James's$, with an average of 3027.27.
Around the average in-strength density values we find wards like
$Woodside$, $Hackney$ $ Central$, $Cantelowes$, etc.. The
geographical distribution for the wards in-strength is given in the
map at the bottom right of Fig.\ref{2}.

As we noticed before for the degree density,  in this case we got
 that the out-strength density values range is just the $11\%$
  of the in-strength density values range. In fact business areas tend to be concentrated in certain zones, defined by
high values of in-strength. The out-strength values reflect the
residential habits: the fact people tend to live in places that are
more widely distributed around the whole city.

 The differences between
out and in vertex properties are better understood if we look at the
experimental density distributions of probability for those
quantities. In Fig.\ref{3} we show the out-degree/strength
distributions. Since they're very similar in shape, we can discuss
 them together. On the left we show the plots on a linear-log
scale. In this way the shapes look very similar to log-normal
distributions, that is the distribution of a measure whose logarithm
is normally distributed. Nevertheless if we look at the same
distributions on a log-linear scale we notice that the tail is a
straight exponential.

\begin{figure}[!htbc]\center
             \includegraphics[width=0.48\textwidth]{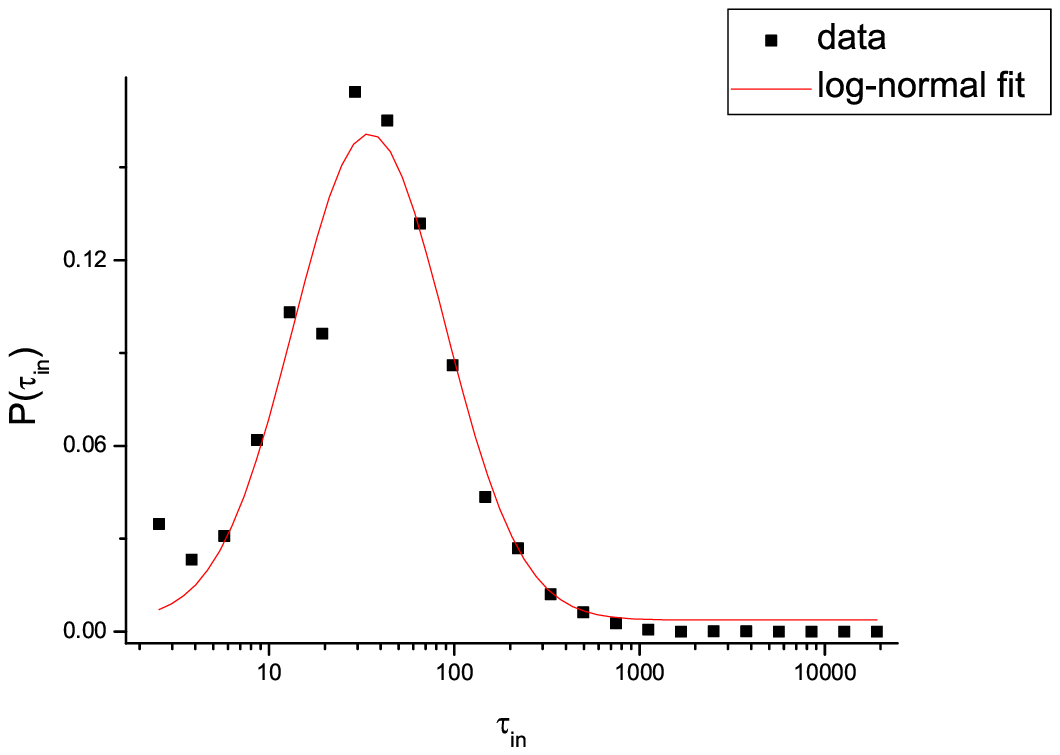}
         \includegraphics[width=0.48\textwidth]{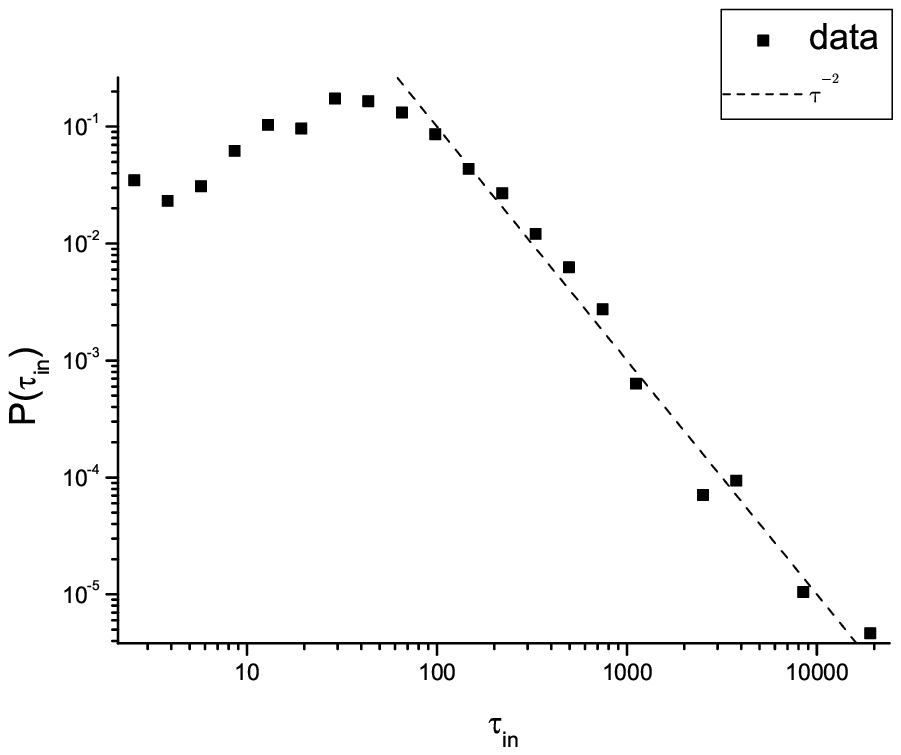}
          \includegraphics[width=0.48\textwidth]{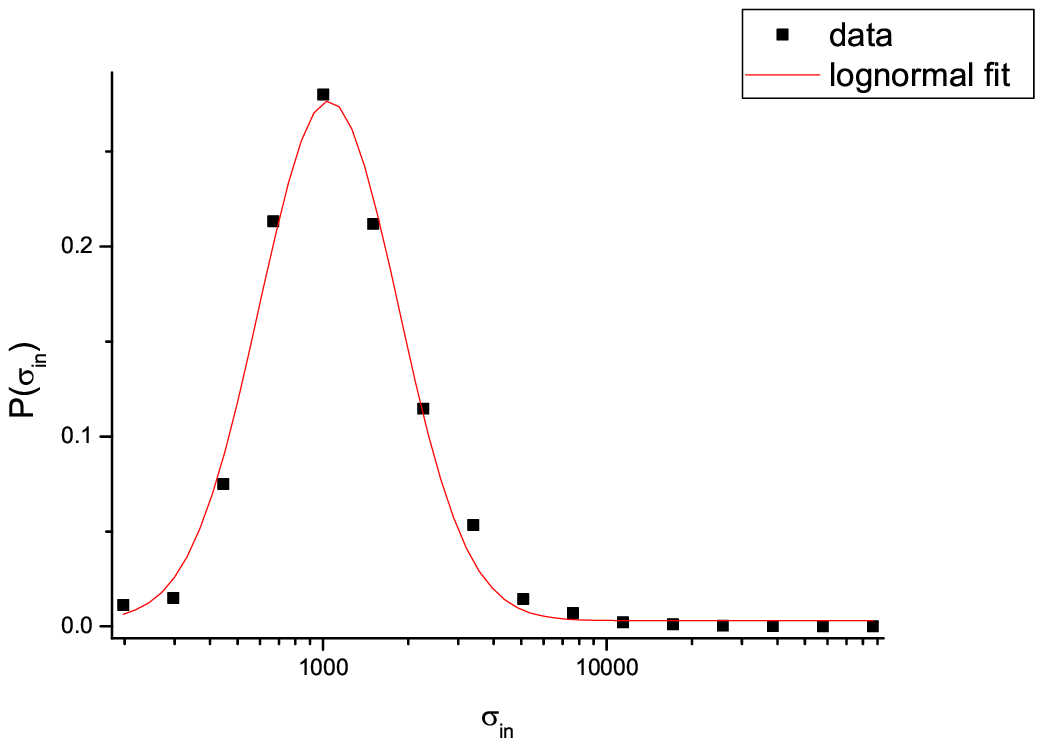}
         \includegraphics[width=0.48\textwidth]{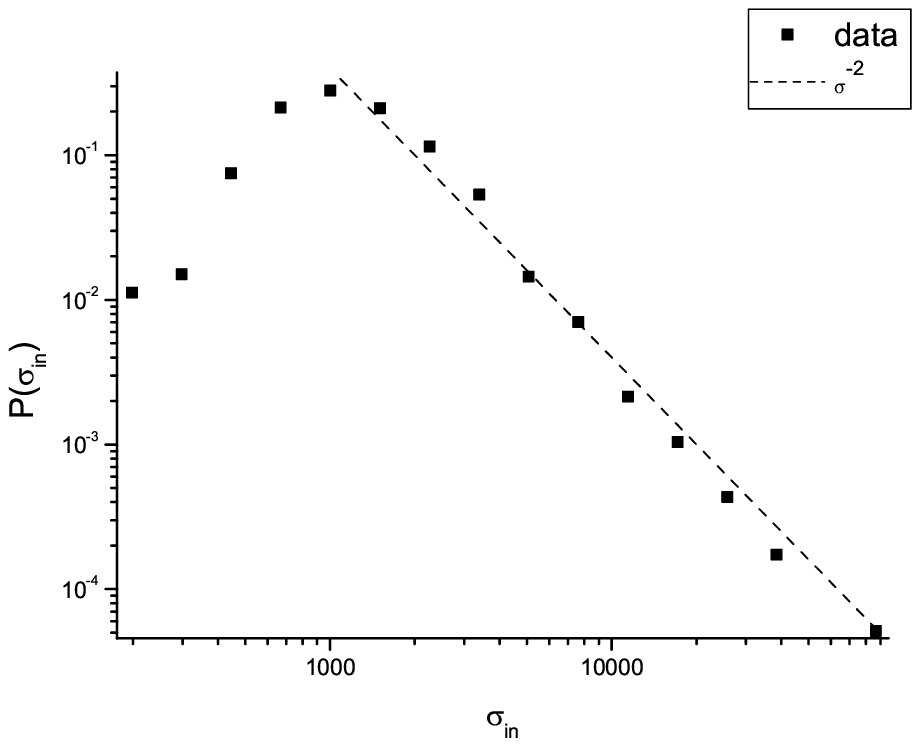}
\caption{\label{4} Top panels: Distribution for the in-degree
density $\tau_{in}$. On the left the log-linear scale shows the
log-normal shape, while the log-log scale on the right shows the
power law behavior of the tale of the distribution. Bottom
panels:Distribution for the in-strength density $\sigma_{in}$. On
the left the log-linear scale shows the log-normal shape, while
the log-log scale on the right shows the power law behavior of
the tale of the distribution. }
\end{figure}

In Fig.\ref{4} we show the in-degree and the in-strength
distributions. As in the previous case the shapes are very similar.
On a linear-log scale we find again the shape of a log-normal
distribution. However the difference with the previous case is that
when we look at the tail of the distribution on a log-log scale, we
see that the distributions fall as a power law with exponent around
$-2$.

In the next section we will give a simple interpretation of those
results.

\section{The Model}

To understand the statistical behaviour of the data, we focus on the
fact that the phenomena we are dealing with, that is the
metropolitan business centers and the metropolitan human residential
settlements, are strictly related and influence each other during
the growth of the city.

To understand the former phenomena, we have to look at the bottom
right panel of Fig.\ref{2}. The center of London, spreading from
$the$ $City$ to $West$ $End$, has the biggest concentration of jobs
in London, going away from this center the business centers
gradually decrease. Then we can notice three other smaller business
centers in the same map, $Heathrow$ in the west, $Croydon$ in the
south and $Isle$ $of$ $Dogs$ east of the center. Analysing the
nearest neighbour properties of those areas,we can see that, on a
smaller scale, they reproduce the behaviour of central London.

The other strong evidence is that the strength and degree
distributions have a peak and a power law tail with exponent around
$-2$. This tail can be explained if we consider a distribution of
points in a circle where the occupation probability $\Pi$ is
proportional to the inverse of the square of the distance from the
center $r$,
\begin{equation}\label{e1}
\Pi^{in}(r)\propto \frac{1}{r^2}.
\end{equation}
 If we define the in-strength $s^{in}$ in this case as the number of points falling in a certain area of the circle, then the in-strength will be
  completely dependent on the
occupation process and we will have that $<s^{in}>\propto
\frac{1}{r^2} $. To calculate the probability density function
$P(s^{in})$ for the in-strength, we can calculate the probability
density function for $r^{-2}$. In general we have that, if
$<s^{in}>\propto r^{-\alpha} $, then
$P(s^{in})=P(r^{-\alpha})\frac{dr^{-\alpha}}{ds^{in}}\propto
P(r)r^{\alpha+1}\frac{dr^{-\alpha}}{ds^{in}}\propto
s_{in}^{-1-2/\alpha}$. It is thus easy to see that, if $\alpha=2$,
$P(s^{in})\propto s_{in}^{-2}$.

\begin{figure}[!htbc]\center
             \includegraphics[width=0.48\textwidth]{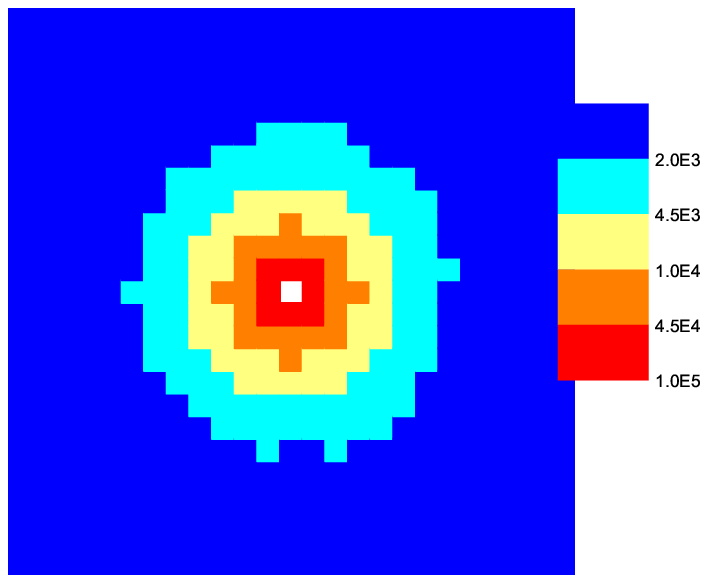}
         \includegraphics[width=0.45\textwidth]{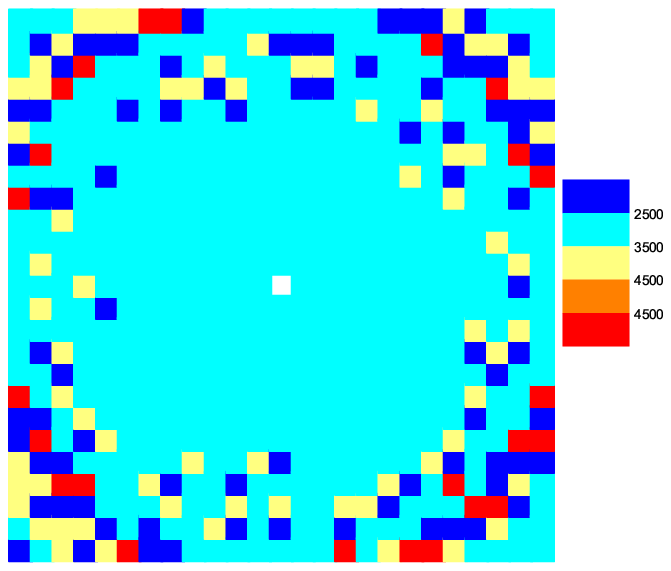}
         \includegraphics[width=0.48\textwidth]{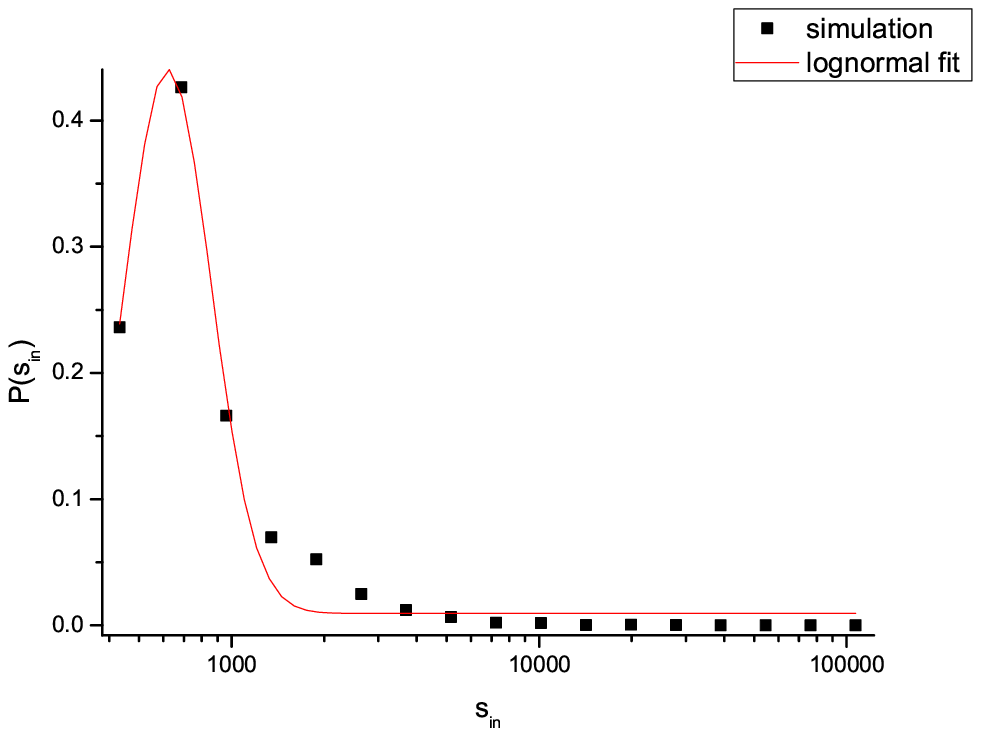}
         \includegraphics[width=0.48\textwidth]{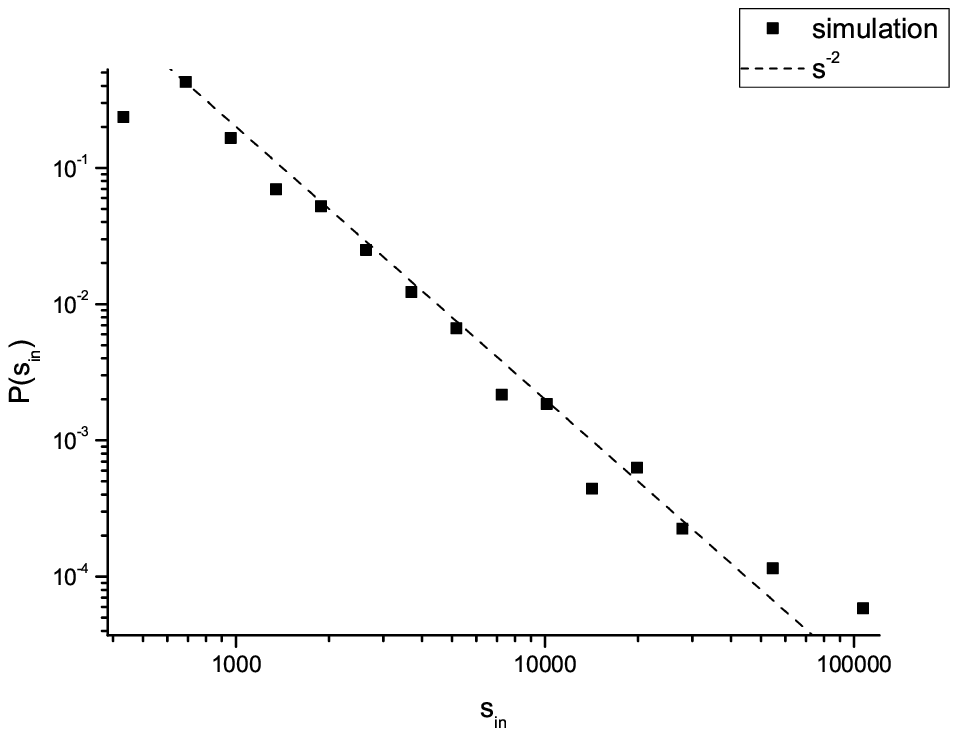}
          \includegraphics[width=0.48\textwidth]{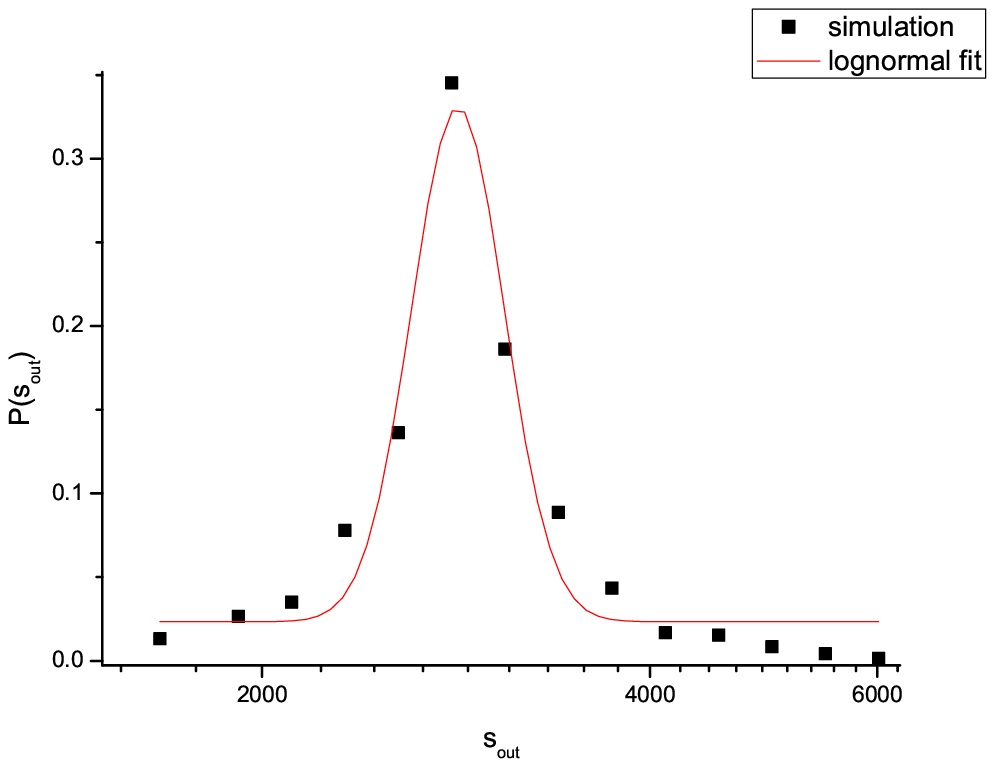}
         \includegraphics[width=0.48\textwidth]{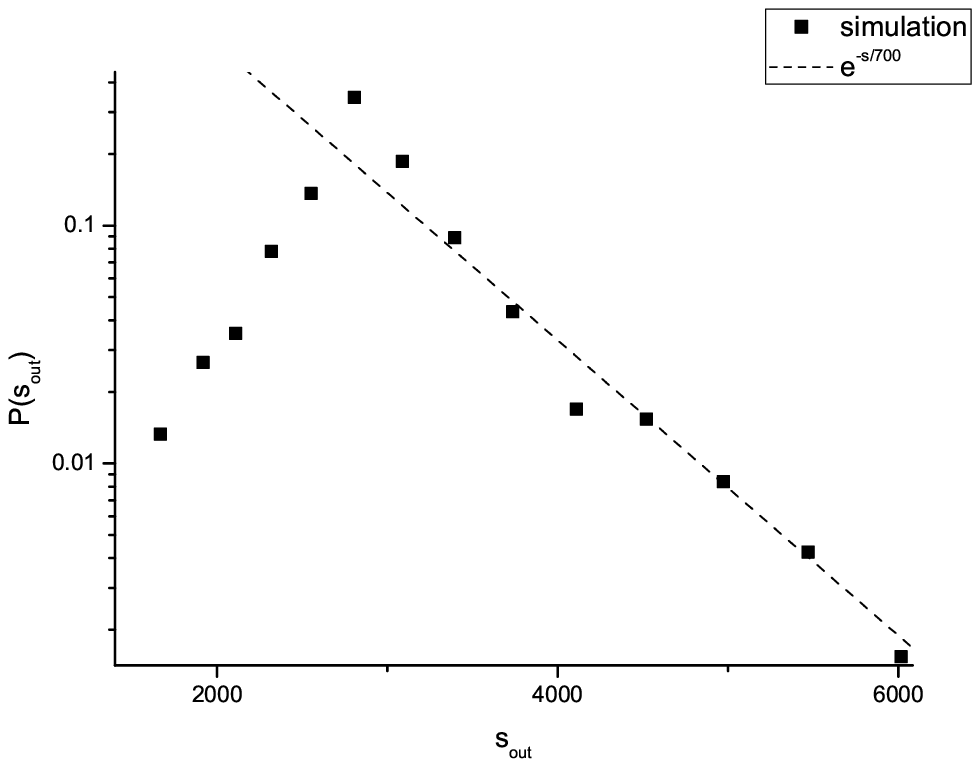}
\caption{\label{5} Square lattice simulation results. Top left panel: simulation for the in-strength, representing
the business centers distribution. Top right: simulation for the out strength, representing the settlement distribution.
Central panels: in-strength distribution: on the left the log-linear scale evidences the
log-normal shape, while the log-log scale on the right evidences the
power law behavior of the tale of the distribution. Bottom
panels: distribution for the out-strength. On
the left the log-linear scale evidences the log-normal shape, while
the linear-log scale on the right evidences the exponential behavior of
the tale of the distribution. }
\end{figure}

The peaked curve can be explained by the asymmetries of the city,
that is London is not circular. To demonstrate this, we performed a
simulation on a square lattice with 625 cells that we populated with
1875000 points with the probability given in Eq.\ref{e1}(these
parameters are chosen to reproduce the London statistics). In the
top left of Fig.\ref{5} we show the resulting map for the in
strength while in the central panels of the same figure the
resulting in-strength distribution. Those results have to be
compared with the distribution in Fig\ref{4}. Although we don't
capture the behaviour of the distribution for the  values of the
in-strength going to zero, we can notice that for the small values
of the in-strength, the curves are very similar.

We can then assume that the in-strength distribution, that is the
distribution of business metropolitan areas, is a geographical
dependent variable. This means that once the business areas are
settled, then they will grow just as an organism does,  trying to be
 as compact as possible and with a radial homogeneous
distribution.

To understand the properties of the out-degree/strength
distributions, that is where people decide to live,  we can notice
(Fig.\ref{2}) that people tend to live close to their  workplace,
but not in the  wards where there is a massive business activity. We
interpret this observation in a stochastic growing model on the
square lattice whose cells represent the wards of London. So, as we
did for the in-strength distribution, we consider a square lattice
with 625 cells. While the business centers are populated with the
probability given in Eq.\ref{e1}, the residential ward $i$ will be
populated with a probability given by:
\begin{equation}\label{e2}
\Pi^{out}_i\propto \frac{1}{s^{in}_ir_i^2},
\end{equation}
where $r_i$ is the Euclidean distance from the ward $i$ to the
center of the lattice. The probability in Eq.\ref{e2} takes into
account the fact that generally people tend to live close to their
workplace with a rate that is  proportional to the inverse of the
square of $r$, but people don't want to live in an area completely
devoted to business, so with a inverse proportional dependence on
the in-strength $s^{in}$ of the ward. The resulting simulated map
for the out-strength is given in the top right panel of Fig.\ref{5}.
From the bottom panels of the same figure we can see that the
 probability distribution obtained for the out-strength possesses the
required features, that is a peaked  distribution with exponential
tail.

\section{Conclusions}

In this work we analysed the network of commuters in London. Our
empirical analysis is in itself  important and unique. The data from
 2001 census regarding the working habits of people of London
are organised and contextualised in the framework of network theory.
The organization of a city relies on many levels of complexity, from
the social differences between people to the geographical
constraints in the landscape of the city itself. Our research
focuses on the organization seen as a result of phenomena related to
the geographical locations of  jobs and the accessibility of those
places. We believe that in order to understand the organization of
the city, those habits are the most important to consider,  and
actually London is the biggest and most productive city of western
Europe.

We show that the power law for the distribution of business centers
can be considered as the result of a pure geographical distribution
of business areas, that is business centers tend to aggregate to
preexisting business centers. In the model we propose the
residential distribution in the city is described as a phenomena
dependent on the distribution of business centers. This dependence
on the business centers is shown to be  $anti-preferential$, that is
people want to be close to their place of work, but don't want to
live in an area devoted to business. The simulations seem to agree
with the real data even if the model is minimal. In fact we showed
how in London, beside the bigger activity center that is in Central
London, other activity centers emerge at different scales. In our
minimal model this effect is not considered, so that it can be seen
as a model of local development that can be used at different
scales.

\begin{acknowledgments}
We thank the European Union Marie Curie Program NET-ACE (contract
number MEST-CT-2004-006724) for financial support. We would also
like to thank Margarethe Theseira and the staff of the Greater
London Authority for their help and support while this work was
completed.

\end{acknowledgments}

\end{document}